\documentclass[conference]{IEEEtran}
\usepackage{cite}
\usepackage{amsmath,amssymb,amsfonts}
\usepackage{algorithmic}
\usepackage{fontawesome5}
\usepackage{graphicx}
\usepackage{textcomp}
\usepackage{multirow}
\def\BibTeX{{\rm B\kern-.05em{\sc i\kern-.025em b}\kern-.08em
    T\kern-.1667em\lower.7ex\hbox{E}\kern-.125emX}}

\usepackage{cite}
\usepackage{amsmath,amssymb,amsfonts}
\usepackage{algorithmic}
\usepackage{graphicx}
\usepackage{tabularx,booktabs,makecell,enumitem}
\setlist[itemize]{nosep,leftmargin=*,label=•}
\usepackage[dvipsnames]{xcolor}
\usepackage[final]{microtype}
\usepackage[italic]{mathastext}
\usepackage{libertine}
\usepackage[T1]{fontenc}
\usepackage{textcomp}
\usepackage{booktabs}
\usepackage[varqu,varl]{zi4}
\usepackage[all]{nowidow}
\usepackage{fancyhdr}

\usepackage{fontawesome5} 

\definecolor{InsightBg}{HTML}{dae3e4}
\definecolor{OptBg}{HTML}{d6ebdd}

\newcommand{\intext}[1]{%
  \raisebox{-0.2em}{\includegraphics[height=1em]{#1}}%
}
\newcounter{insightcounter}
\newcommand{\Insight}[1]{%
  \refstepcounter{insightcounter}%
  \vspace{1ex}%
  \par\noindent%
  \colorbox{InsightBg}{%
    \parbox{\dimexpr\linewidth-2\fboxsep}{%
      \textbf{\intext{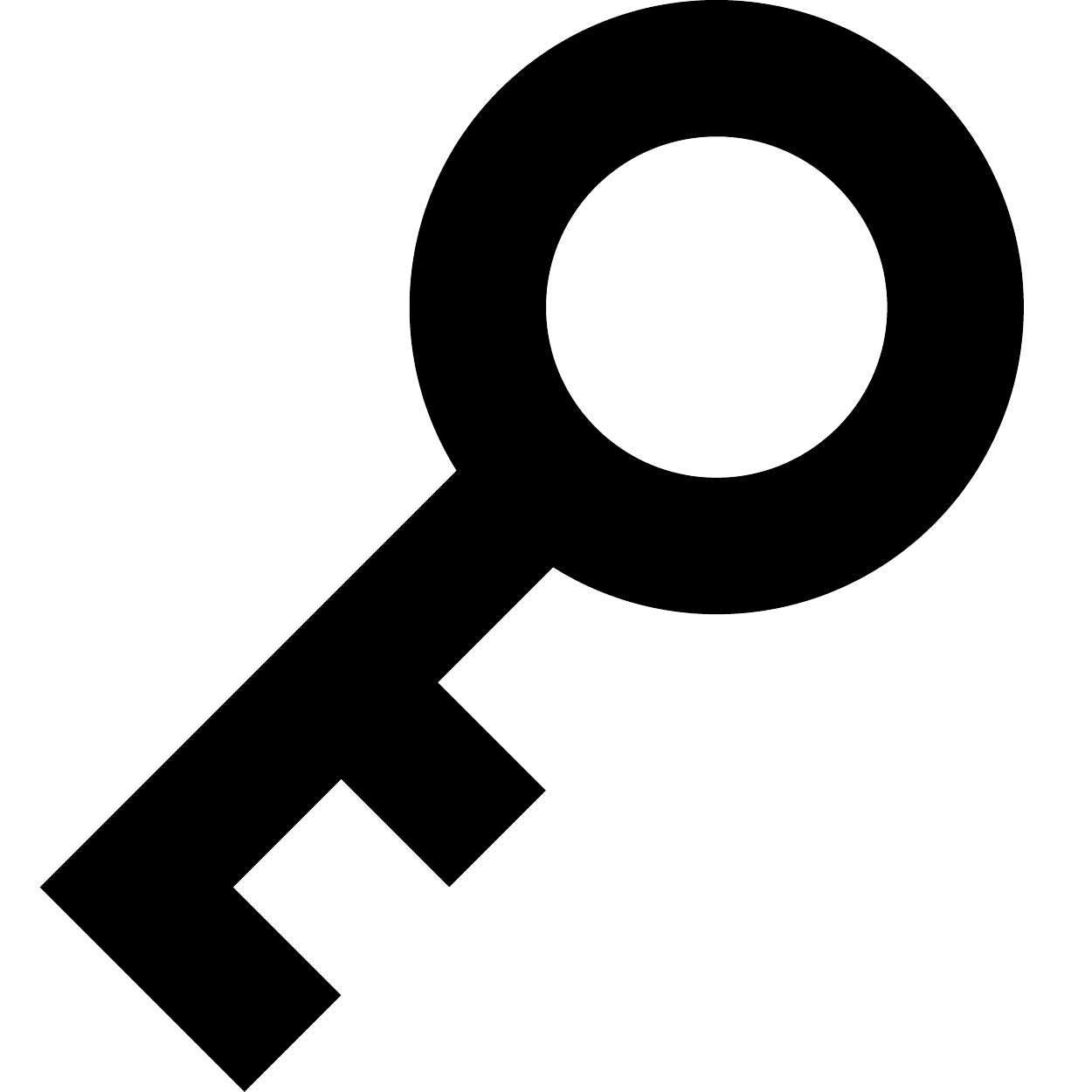} Insight~\theinsightcounter:}~#1%
    }%
  }%
  \par
  \vspace{1ex}
}

\newcounter{optcounter}

\usepackage{cite}
\usepackage{todonotes}
\usepackage{amsmath,amssymb,amsfonts}
\usepackage{algorithmic}
\usepackage{graphicx}
\usepackage{textcomp}
\usepackage{url}
\usepackage{stfloats}
\def\BibTeX{{\rm B\kern-.05em{\sc i\kern-.025em b}\kern-.08em
    T\kern-.1667em\lower.7ex\hbox{E}\kern-.125emX}}

\begin{document}

\title{Confidential LLM Inference: Performance and Cost\\ Across CPU and GPU TEEs
}

\author{\IEEEauthorblockN{Marcin Chrapek}
\IEEEauthorblockA{
\textit{ETH Zurich}\\
Zurich, Switzerland \\
marcin.chrapek@inf.ethz.ch}
\and
\IEEEauthorblockN{Marcin Copik}
\IEEEauthorblockA{
\textit{ETH Zurich}\\
Zurich, Switzerland}
\and
\IEEEauthorblockN{Etienne Mettaz}
\IEEEauthorblockA{
\textit{ETH Zurich}\\
Zurich, Switzerland}
\and
\IEEEauthorblockN{Torsten Hoefler}
\IEEEauthorblockA{
\textit{ETH Zurich}\\
Zurich, Switzerland}
}

\maketitle

\begin{abstract}
Large Language Models (LLMs) are increasingly deployed on converged Cloud and High-Performance Computing (HPC) infrastructure.
However, as LLMs handle confidential inputs and are fine-tuned on costly, proprietary datasets, their heightened security requirements slow adoption in privacy-sensitive sectors such as healthcare and finance.
%
%
We investigate methods to address this gap and propose Trusted Execution Environments (TEEs) as a solution for securing end-to-end LLM inference.
We validate their practicality by evaluating these compute-intensive workloads entirely within CPU and GPU TEEs.
On the CPU side, we conduct an in-depth study running full Llama2 inference pipelines (7B, 13B, 70B) inside Intel's TDX and SGX, accelerated by Advanced Matrix Extensions (AMX). 
We derive 12 insights, including that across various data types, batch sizes, and input lengths, CPU TEEs impose under 10\% throughput and 20\% latency overheads, further reduced by AMX.
We run LLM inference on NVIDIA H100 Confidential Compute GPUs, contextualizing our CPU findings and observing throughput penalties of 4–8\% that diminish as batch and input sizes grow.
By comparing performance, cost, and security trade-offs, we show how CPU TEEs can be more cost-effective or secure than their GPU counterparts.
To our knowledge, our work is the first to comprehensively demonstrate the performance and practicality of modern TEEs across both CPUs and GPUs for enabling confidential LLMs (cLLMs).

\end{abstract}


\section{Introduction}

Large Language Models (LLMs) dominate the machine learning (ML) landscape~\cite{awaisFoundationalModelsDefining2023, zhaoSurveyLargeLanguage2023}. Exemplified by model families such as GPT~\cite{openaiGPT4TechnicalReport2023a, brownLanguageModelsAre2020} and Llama~\cite{touvronLLaMAOpenEfficient2023, touvronLlamaOpenFoundation2023, Llama4Herd, grattafioriLlama3Herd2024}, they have become prevalent in industry and everyday life across a growing number of domains. LLMs achieve human-like capabilities on multimodal data~\cite{wuNExTGPTAnytoAnyMultimodal2023} and have been applied to disciplines relying on \textit{confidential} user information, including healthcare~\cite{sallamChatGPTUtilityHealthcare2023}, finance~\cite{wuBloombergGPTLargeLanguage2023}, sentiment analysis~\cite{araciFinBERTFinancialSentiment2019}, legal cases~\cite{cuiChatLawOpenSourceLegal2023}, and document translation~\cite{kocmiLargeLanguageModels2023}. Simultaneously, the ever-increasing size of LLMs has led to changes in their deployment strategies. LLMs ranging from billions to trillions of parameters necessitate state-of-the-art hardware to meet their performance demands, which is frequently provided by cloud service providers (CSPs).

\begin{figure}[!t]
  \centering
  \includegraphics[width=\linewidth]{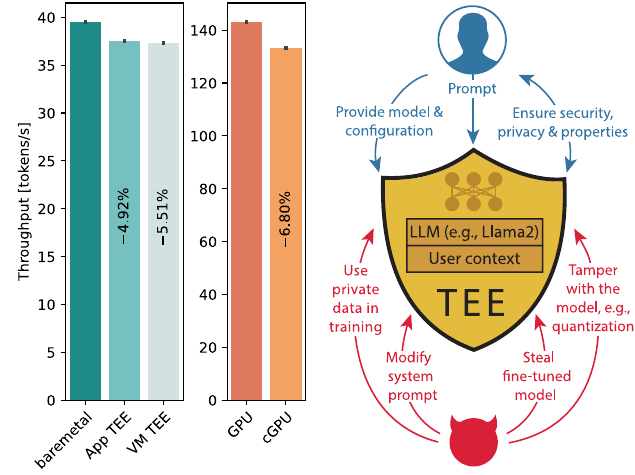}
  \caption{Example attacks on LLMs that TEEs protect against and our performance results for Llama2 7B inference in two CPU TEEs: a Virtual Machine (VM, TDX) and an application-based (App, SGX) one, and a GPU TEE (cGPU).}
  \label{fig:child_poster}
  \vspace{-1em}
\end{figure}

However, deploying within the cloud carries security risks for LLMs that operate on expensive and confidential data. Figure~\ref{fig:child_poster} shows attacks that cloud providers, cluster administrators, and other tenants can carry out to leak model information and influence inference results. Data confidentiality and intellectual property (IP) theft pose critical threats to LLMs, for which the cost of engineering and obtaining datasets is substantial. With training and fine-tuning alone amounting to tens of millions of dollars~\cite{sharirCostTrainingNLP2020}, any security breach involving LLMs is increasingly costly for CSPs, model providers (e.g., MetaAI, OpenAI, financial or healthcare institutions), and end users. 



While such threats might seem distant, they led to companies banning internal LLM use~\cite{SamsungBansGenerative2023} and are tangible. For example, health records processed by a cloud-deployed LLM for insurance could be stolen and used maliciously. Even if leveraged solely for illicitly training another model, such a model might then be probed through public queries, reconstructing sensitive data, including names, addresses, Social Security numbers, and full medical histories~\cite{carliniExtractingTrainingData2021,patilCanSensitiveInformation2023}. We also observe reports of backlash against leveraging user data for AI features~\cite{ngAdobeSaysIt} as more companies offer personalized AI (e.g., Meta's AI Studio or Adobe Creative). 

The security community has addressed the issues of ML IP theft and data confidentiality by employing three primary approaches: model modifications (e.g., watermarking and user authentication~\cite {xueIntellectualPropertyProtection2022, knottCrypTenSecureMultiParty2022, dowlinCryptoNetsApplyingNeural}), cryptographic methods (e.g., homomorphic encryption~\cite{ebelOrionFullyHomomorphic2025}), and trusted execution environments~\cite{moMachineLearningConfidential2023}. We conduct an analysis of these techniques in Section~\ref{sec:related_work} and show that TEEs currently provide the only viable method for protecting LLM inference. TEEs offer a practical balance between robust security properties, performance costs, and generalizability. 

%




Our work focuses on quantifying usability and performance overheads of TEEs for protecting LLM inference by evaluating representative implementations of both CPU and GPU TEEs.
In Section~\ref{sec:hardship}, we start by evaluating the CPU side and conducting an in-depth study of Intel's Trust Domain Extensions (TDX) and Software Guard Extensions (SGX), representing common approaches to implementing TEEs: through virtual machines (VMs) and processes. 
We identify the best-performing frameworks and present the TEE performance overheads for throughput and latency in an end-to-end Llama2 (7B, 13B, and 70B) inference pipeline across various batch sizes, input lengths, and data types.
Leveraging this compute-intensive workload, we derive 12 key insights on the performance of confidential LLM (cLLM) hosting, with practical guidelines for users and cloud providers.
Our insights can be generalized to other TEE deployments and LLM systems. 
For example, we demonstrate how Advanced Matrix Extensions (AMX) directly result in lower overheads for TEEs.
%
Figure~\ref{fig:child_poster} displays our example performance results, showing that TEEs for LLMs incur only 4-7\% throughput reduction compared to overheads of up to 100s of percent for other applications~\cite{akramPerformanceAnalysisScientific2021, misonoConfidentialVMsExplained2024, coppolinoExperimentalEvaluationTEE2025}. 

In Section~\ref {sec:gpu_perf}, we present GPU results evaluated on NVIDIA's H100s that put our CPU results in perspective.
We compare these two setups, considering cost, performance, and security.
For example, we show that with AMX, CPU-based TEEs can be more cost-efficient than confidential NVIDIA H100 GPUs.
%
%
%
%
%
Finally, in Section~\ref{sec:RAG}, we evaluate one of the most common LLM extensions: Retrieval Augmented Generation (RAG)~\cite{gaoRetrievalAugmentedGenerationLarge2024}.
We run full RAG pipelines, including Elasticsearch databases, in TEEs, and report their 7\% overheads.
We demonstrate how our lessons on CPU and GPU TEEs directly extend to these types of deployments.
%
To our knowledge, our research is the first to comprehensively demonstrate the performance and practicality of modern TEEs across both CPUs and GPUs for enabling cLLMs.
Our work can be replicated and evaluated seamlessly on other systems with the open-source implementation and configuration we release.
\textbf{In summary, our contributions are:}
\begin{enumerate}
    \item Demonstrating how TEEs currently offer the only pragmatic solution for protecting LLM inference.
    \item Characterizing performance of CPU TEEs (SGX, TDX) on Llama2 (7B/13B/70B), showing overheads of less than 10\% for throughput and 20\% for latency, identifying sources of performance degradation and optimal configurations.
    \item Demonstrating how these relate to GPU TEEs by comparing with CPU TEEs in terms of cost-effectiveness, performance, and security.
    \item Open-sourcing our configuration\footnote{\url{github.com/spcl/confidential-llms-in-tees}} and drawing 12 insights from empirical results, guiding efficient deployment and TEE system design.
\end{enumerate}

\section{Protection Mechanisms for LLM Inference}
\label{sec:related_work}

Three approaches can be used to protect LLM inference: machine learning (ML) methods, cryptographic methods such as Homomorphic Encryption (HE) and multiparty computation (MPC), and Confidential Computing (CC)~\cite{mulliganConfidentialComputingBrave2021}. 

\textbf{ML methods:} As noted in literature~\cite{xueIntellectualPropertyProtection2022}, current ML methods focus on post hoc detection of intellectual property (IP) theft in the form of model verification and passive protections, falling short in actively covering against model or data theft. Example approaches include using signatures embedded in the model with model theft verification using input/output pairs~\cite{laoDeepAuthDNNAuthentication2022}, passport~\cite{fanRethinkingDeepNeural2019} or backdoor~\cite{xueActiveDNNIP2020} authentication, and watermarks in model output or weights used for ownership verification~\cite{szyllerDAWNDynamicAdversarial2021, boenischSystematicReviewModel2021}. 

While these protect against specific attacks, they do not provide exhaustive and measurable security properties. The cost of losing confidentiality or IP theft makes it challenging to rely only on them. Additionally, ML methods frequently require expensive retraining, alter the model's accuracy, fail to secure the confidentiality of user prompts~\cite{xueIntellectualPropertyProtection2022}, and cannot be combined together~\cite{szyllerConflictingInteractionsProtection2023}. Cryptographic approaches, such as HE and MPC, address these issues through strong cryptographic protocols.

\textbf{Cryptographic methods:} HE allows conducting mathematical and logical operations on encrypted data without decrypting~\cite{acarSurveyHomomorphicEncryption2018}. HE has been explored in the context of DNNs~\cite{dowlinCryptoNetsApplyingNeural, leePrivacyPreservingMachineLearning2022, woodHomomorphicEncryptionMachine2020}. However, except for a few structured examples~\cite{chrapekHEARHomomorphicallyEncrypted2023, burkhalterTimeCryptEncryptedData2020}, the state-of-the-art HE is not practical. HE operations on encrypted data can have up to 10,000x performance and size overheads, taking minutes to conduct simple MNIST~\cite{dowlinCryptoNetsApplyingNeural} or RESNET~\cite{ebelOrionFullyHomomorphic2025} inference, and making LLM inference intangible. HE approaches also do not provide integrity protection. MPC is close to HE and has similar practicality issues, but involves multiple parties~\cite{viandMarbleMakingFully2018a}.

\textbf{Confidential Computing:} CC offers an alternative in the form of TEEs, using security primitives implemented in hardened hardware. Compared to HE and MPC, which rely on obscuring data and functions, TEEs offer a secure and isolated environment, frequently referred to as an \textit{enclave}. Users can verify enclaves in a safe, hardware-enabled process called \textit{attestation}. TEEs ensure the confidentiality and integrity of running programs and their data, protecting against external and privileged attackers, such as system administrators. TEEs achieve this by prohibiting access to or modification of the memory contents of running programs~\cite{sabtTrustedExecutionEnvironment2015}, including sensitive data like weights or user-confidential information.
TEEs widely available on CSP platforms include CPU-based examples such as AMD's Secure Encrypted Virtualization-Secure Nested Paging (SEV-SNP)~\cite{kaplanAMDSEVSNPStrengthening}, Intel's SGX~\cite{mckeenInnovativeInstructionsSoftware2013, hoekstraUsingInnovativeInstructions2013, costanIntelSGXExplained2016} and TDX~\cite{chengIntelTDXDemystified2024}, ARM's TrustZone~\cite{pintoDemystifyingArmTrustZone2019} and CCA~\cite{liDesignVerificationArm2022}, and GPU-based examples such as NVIDIA's Confidential GPUs~\cite{nertneyConfidentialComputeNVIDIA2023}.

Although TEEs do not provide the formal guarantees of HE or MPC, they still offer quantifiable defenses, particularly against integrity attacks that HE and MPC cannot address. Unlike many ML approaches, TEE protection mechanisms actively ensure enforcement of trust boundaries. However, performance and programmability are often cited as the two primary limitations of TEEs~\cite{akramPerformanceAnalysisScientific2021}. Because their security primitives lie on the critical path, TEEs incur non-negligible overhead. Nonetheless, as we show in our evaluation, TEE's overheads remain substantially lower than those imposed by HE schemes. Similarly, although TEEs require some security expertise, leveraging VM TEEs and frameworks like Gramine~\cite{tsaiGrapheneSGXPracticalLibrary2017} eliminates the need for application modifications necessary for HE and ML methods.


\Insight{TEEs offer a practical balance between security, performance, and programmability.}

\section{CPU TEEs}
To investigate practical deployments, we limit ourselves to CPU TEEs offered by major CSPs. The options are limited to AMD and Intel since other TEEs, such as ones based on RISC-V~\cite{leeKeystoneOpenFramework2020} or ARM~\cite{pintoDemystifyingArmTrustZone2019}, are not widely available. We selected Intel's TEEs for two reasons. Firstly, they include support for AMX, an on-chip matrix operation hardware accelerator that introduces CPU-native support for formats such as brain-floating-points (\verb|bfloat16|) and 8-bit integers (\verb|int8|). AMX improves LLM performance 2-6x~\cite{naUnderstandingPerformanceImplications2024}~(Figure~\ref{fig:AMX_scaling}), and we investigated whether these units also impact the performance of TEEs~(Section~\ref{sec:AMX}). Secondly, they provide us with two common ways of implementing TEEs (VMs and processes) within the same system, covering other TEEs and enabling an apples-to-apples comparison without scaling performance results. For example, AMD’s TEE stack relies on similar security mechanisms to Intel's TDX, resulting in close benchmark overheads~\cite{misonoConfidentialVMsExplained2024}. 

\label{sec:hardship}




\subsection{Process-based TEEs: SGX}
\label{sec:SGX}

SGX programming model differentiates between SGX-protected and unprotected program sections. The former, located within an enclave, is safeguarded by SGX capabilities, while the latter is unsecured. SGX has two sources of overhead. First, data in the enclave is protected by memory encryption and integrity checks. Second, operations switching to the SGX unprotected program sections (e.g., IO such as reading a file) save SGX state and invalidate the caches.

SGX enclaves are frequently deployed on top of library operating systems (OSs) created for TEEs, such as Gramine~\cite{tsaiGrapheneSGXPracticalLibrary2017} or Occlum~\cite{shenOcclumSecureEfficient2020}. These are lightweight layers between the host system and applications, intercepting any system calls to ensure they are conducted securely. These address some inconveniences of the original SGX software development kit (SDK), which required users to manually rewrite applications with secure and insecure sections. 

\begin{figure}[]
  \centering
  \includegraphics[width=\linewidth]{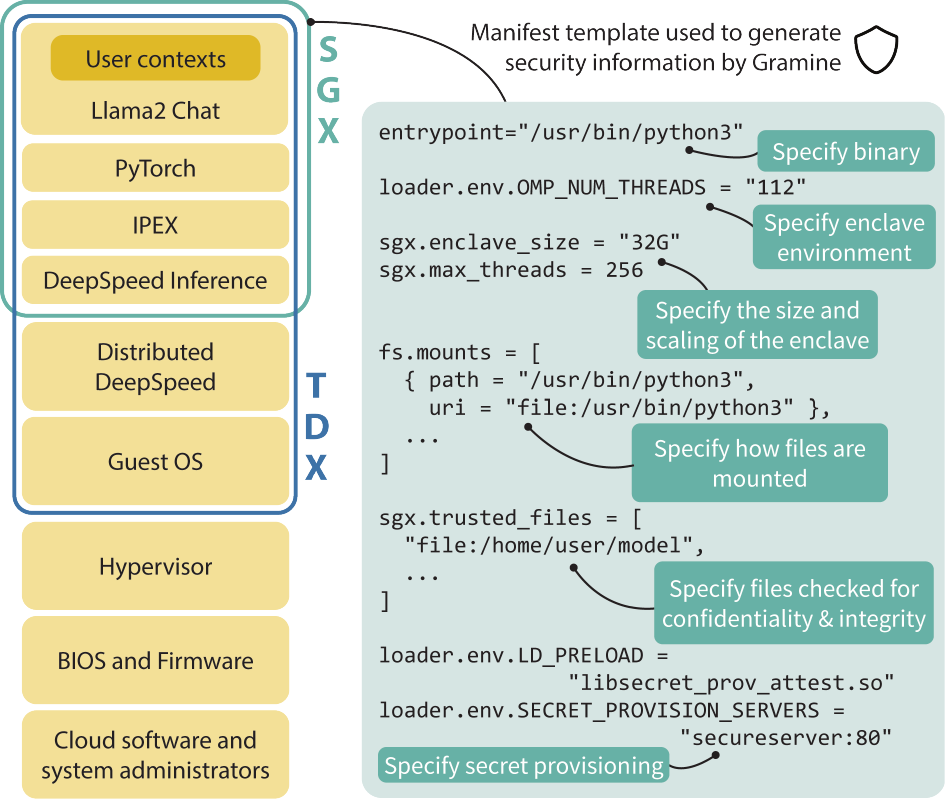}
  \caption{Our software stack with the layers we protect in Intel TDX and SGX, and an extract from our Gramine manifest template file.}
  \label{fig:TEEs}
  \vspace{-1em}
\end{figure}

In our study, we use the open-source Gramine~\cite{tsaiGrapheneSGXPracticalLibrary2017} library OS that 
enables porting applications to SGX without significant code modifications. Unlike alternatives, it has lower requirements on the format of protected applications~\cite{shenOcclumSecureEfficient2020}, is not proprietary~\cite{arnautovSCONESecureLinux2016}, and is mature~\cite{segarraServerlessConfidentialContainers2024}. Gramine automatically applies integrity and confidentiality protections to storage, simplifies attestation, and transparently uses instructions for leaving and entering the SGX enclave during system calls. To increase performance, Gramine emulates some system calls without exiting the SGX enclave. However, if a given call is not implemented fully, it can result in considerable overhead. As we experienced firsthand, this can create a challenge while working with SGX, especially with complex workloads.

Gramine exposes its features via a Manifest file, which outlines the enclave size, the number of threads, the binary to be run, the files that can be trusted, and where to obtain the cryptographic decryption keys. Figure~\ref{fig:TEEs} shows an example excerpt from a Manifest file.


 
\subsection{VM-based TEEs: TDX}
TDX is a virtual machine (VM) based TEE that introduces security features using a hardened hardware-enabled kernel virtual machine (KVM) hypervisor. In the TDX security model, the entire VM is protected. This approach aligns well with the CSP virtualization trend and significantly simplifies development, eliminating the need for special functions when entering or exiting the enclave. 
TDX also runs programs within a standard Linux OS, such as Ubuntu, allowing for the easy execution of complex distributed AI frameworks, such as DeepSpeed~\cite{aminabadiDeepSpeedInferenceEnabling2022}, which we use. However, this convenience comes at the price of an increased attack surface. TDX requires trusting the entire VM OS and associated services, rather than just a minimal library OS, like in SGX. Using VMs also implies a virtualization performance tax, which can reach SGX's overheads as we demonstrate in Section~\ref{sec:single_socket}. Furthermore, some security aspects handled by frameworks, such as Gramine, are not performed automatically in TDX. For example, users must protect the filesystem, e.g., by using Linux Unified Key Setup (LUKS)~\cite{fruhwirth2011luks} full-disk encryption. 

To use TDX, one must define a VM with a Quick Emulator (QEMU) command or a libvirt definition file. These specify hardware details, such as boot files, virtual-to-physical core mapping, and memory size, and result in a greater performance impact than enabling TDX (Section~\ref{sec:optimizations}). 

\Insight{TDX is considerably easier to work with than SGX, especially for complex workloads.}


\subsection{Experimental setup}
\label{sec:experimental_setup}
\subsubsection{Hardware and software}
We used two Emerald Rapid dual-socket Intel systems. First EMR1, a dual socket Intel Xeon\textsuperscript{\textregistered}\ Gold 6530 (\$2130~\cite{IntelXeonGold}), each with 32 cores, 16x32GiB 4800MHz DDR5 memory, Ubuntu 23.10, Python 3.10.12, PyTorch 2.2.0, transformers 4.35.2, Intel extension for PyTorch (IPEX) 2.2.0, and oneCCL PyTorch bindings 2.2.0. Second EMR2, a dual socket Intel Xeon\textsuperscript{\textregistered}\ Platinum 8580 (\$10710~\cite{IntelXeonPlatinum}), each with 60 cores, 16x32GiB 4800MHz DDR5 memory, Ubuntu 24.04, Python 3.10.16, PyTorch 2.3.0, transformers 4.38.1, IPEX 2.3.100, and oneCCL PyTorch bindings 2.3.0. 

\subsubsection{Microbenchmark to select framework}
\begin{figure}[]
  \centering
  \includegraphics[width=\linewidth]{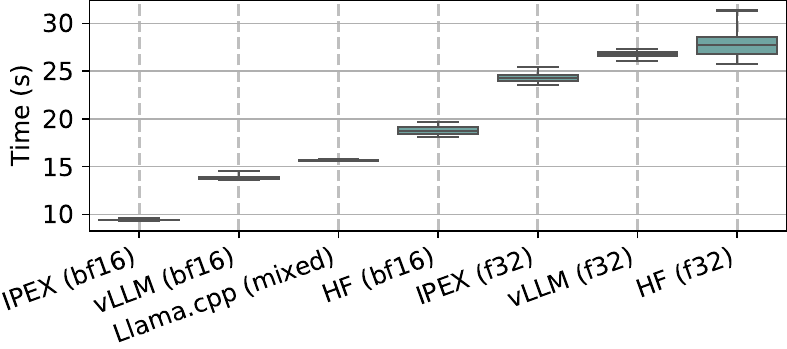}
  \caption{Comparison of single-socket, bare metal wall CPU runtime on EMR1 of different backends and datatypes for Llama2 7B inference over 1024 input and 128 output tokens with beam and batch sizes equal to 1. HF is Hugging Face, bf16 is bfloat16, f32 is float32.}
  \label{fig:frameworks}
  \vspace{-1em}
\end{figure}

To determine the best framework for inference on the CPU, we evaluated multiple popular options and assessed their performance across various data types using an example Llama2 7B LLM. We compared Hugging Face's transformers~\cite{wolf-etal-2020-transformers} (float32, bfloat16), vLLM~\cite{kwon2023efficient} (float32, bfloat16), IPEX, and Llama.cpp~\cite{GgmlorgLlamacpp2025} (mixed datatype). As Figure~\ref{fig:frameworks} shows, IPEX is considerably faster than all other frameworks, with the second vLLM being 50\% slower and Hugging Face 100\% slower. IPEX leverages AMX and its native \verb|bfloat16| support to achieve the best performance. It also utilizes the oneAPI Collective Communications Library (oneCCL), which is fine-tuned for Intel's processors, making it a suitable choice for running across multiple NUMA domains.

\Insight{Leveraging IPEX, and its AMX and oneCCL backends can double CPU inference performance.}

\begin{figure*}[t]
  \centering
  \includegraphics[width=\linewidth]{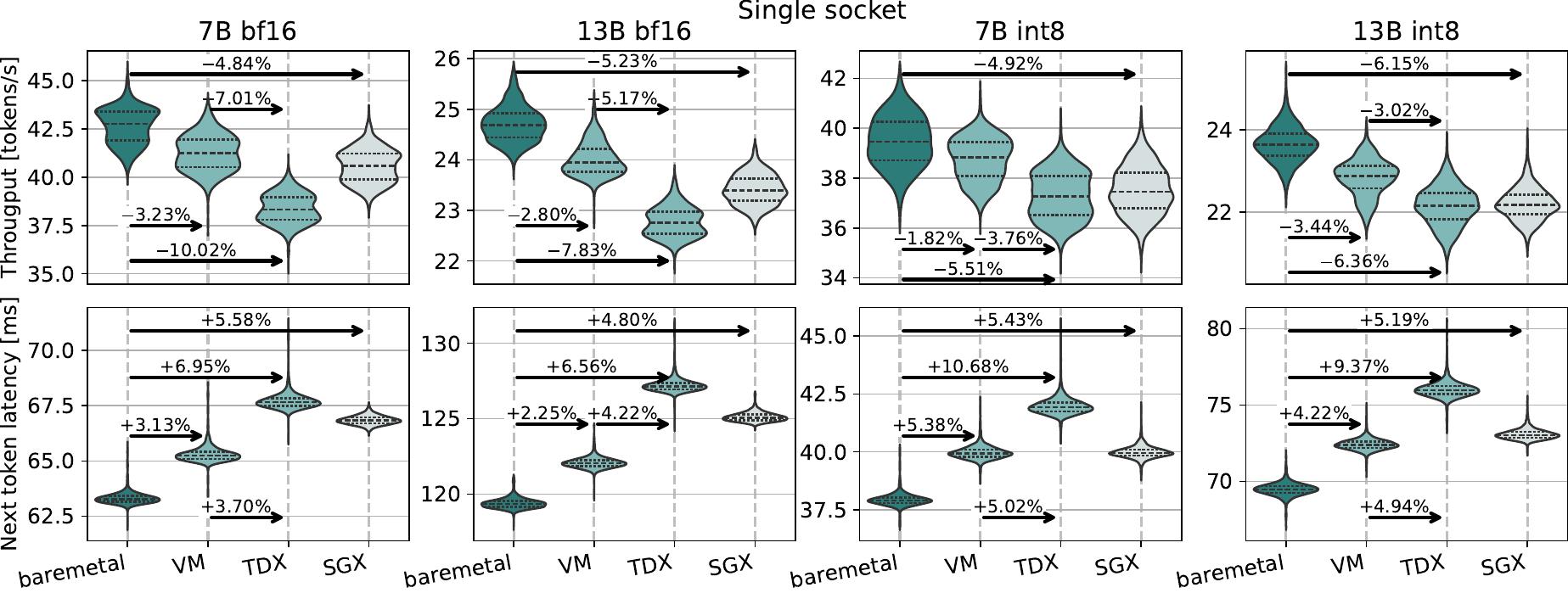}
  \caption{TDX and SGX throughput and latency overheads stay within 4-10\% for Llama2, and 1024 input, 128 output tokens on EMR1. A larger batch size implies increased latency and throughput as less data movement is required per token. Inputs batched are computed on each layer, and a combined result is forwarded to the next layer. Each layer has an increased latency over a single input but a decreased one over $N$ separate inputs (increased throughput).}
  \label{fig:single_NUMA}
  \vspace{-1em}
\end{figure*}

\begin{figure}[t]
  \centering
  \includegraphics[width=\linewidth]{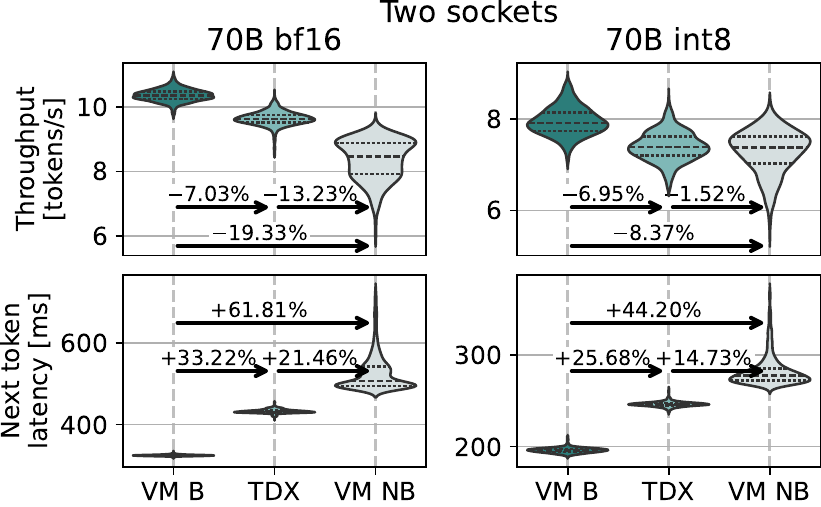}
  \caption{The latency and throughput overheads of TDX over the VM backed with 2MB transparent huge pages (VM B) and VM backed with the same huge pages but without any NUMA binding (VM NB) on EMR1.}
  \label{fig:70B}
  \vspace{-1em}
\end{figure}

\label{sec:performance}
\subsubsection{Experiment details}
We selected Llama2~\cite{touvronLlamaOpenFoundation2023} as a representative example of dense transformers. While subsequent iterations of the Llama family~\cite{Llama4Herd, grattafioriLlama3Herd2024} introduce models of different sizes, larger context windows, or mixtures of experts, they are fundamentally based on the same computational patterns. In this sense, Llama2 also represents well other dense transformer LLMs, such as GPT or OPT. This has been confirmed empirically by consistent performance patterns between these LLMs~\cite{naUnderstandingPerformanceImplications2024}. To verify that this is similar for TEEs, we also evaluated Llama3 8B, GPT-J 6B, Falcon 7B, Baichuan2 7B, and Qwen 7B, and found 3.1-13.1\% overheads, in line with our Llama 7B results.
%
We report user-perceived performance: throughput (tokens per second) and latency (time to receive next token). For latency, we measured the generation time for each token and its inverse for throughput. We run multiple generations for each experiment, measuring at least 1000 output tokens. We used two inference data types: \verb|bfloat16| and \verb|int8|. For the latter, we quantized the models. We evaluate four hardware configurations: the baseline represents results from a bare-metal machine, SGX from Gramine v1.7 running on SGX, VM from a raw VM without security features, and TDX from a TDX-enabled VM.

\subsection{Single socket}
\label{sec:single_socket}
We first establish baseline performance. 
Figure~\ref{fig:single_NUMA} shows the throughput (batch size = 6, beam size = 4) and the next token latency (batch size = 1, beam size = 1). The overhead of Gramine-SGX is between 4.80-6.15\% while for TDX it is between 5.51-10.68\%. TDX adds overhead of 3.02-7.01\% over VM. The results for different data types show that \verb|int8| generally achieves similar throughput to \verb|bfloat16| but almost half the latency. While in SGX, the overheads for \verb|int8| are similar to those for \verb|bfloat16|, TDX shows considerable differences, where \verb|int8| results are better in terms of throughput but worse in terms of latency. For throughput, lower memory movement due to the inference state in \verb|int8| and the corresponding reduction in necessary address translations from guest to host memory results in lower overheads. For latency, memory access costs due to address translations and TEE memory protections are more pronounced when it is lower. All systems have a latency considerably below the average human reading speed of 200 ms/word (approximately 300 words per minute)~\cite{raynerMuchReadLittle2016}, which forms a performance standard that LLMs should meet. As we plot per-token statistics, we noticed outliers for SGX and TDX, which we excluded in the violin plots using a \verb|Z-score| $>$ 3 ($\approx$0.64\% of samples). As visible in the later plots, these do not contribute to the discussion but create considerable noise due to variability in memory encryption.

\Insight{TDX and SGX have overheads as low as 4-10\% for cLLM inference, preserving acceptable service performance.}

The performance of SGX lies between that of a VM and TDX. In our deployment, SGX runs on bare metal, where the host OS has more privileges than a VM and exposes the hardware more directly. TDX, on the other hand, does not have direct access to specific hardware features and must access the underlying system through virtualization layers, such as guest address translations not present in SGX. The results quantify this virtualization tax by showing that running in a VM has an overhead of 1.82-5.38\%. The cost of security is similar for SGX and TDX, as the overheads of SGX over bare metal and TDX over VM are comparable.

\Insight{Compared to SGX, TDX simplifies deployment but increases the trust boundary and pays a virtualization tax of 1-5\%, making SGX more performant.}


\begin{figure*}[t]
  \centering
  \includegraphics[width=\linewidth]{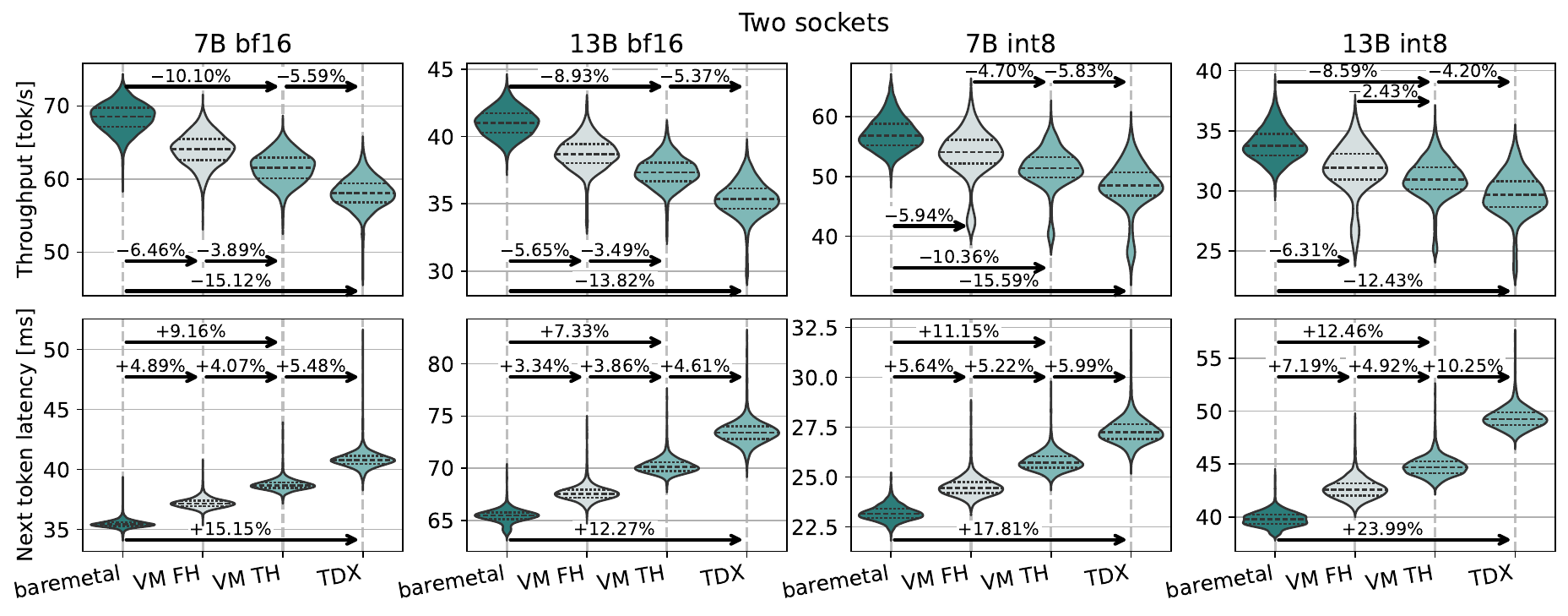}
  \caption{The throughput and latency overheads for VM with full 1GB huge pages (VM FH), 2MB transparent huge pages (VM TH), and TDX on EMR1. The overheads of TDX over VM TH remain at 4-10\%.}
  \label{fig:two_NUMA}
  \vspace{-1em}
\end{figure*}

\section{Tuning CPU TEE overheads}
\label{sec:optimizations}
Our investigation revealed three key areas to achieving acceptable performance within TEEs: appropriate TEE configuration, use of AMX, and optimizing memory efficiency.

\subsection{Configuring TEEs to avoid performance traps}


For SGX, we used the largest possible enclave page cache (EPC), which significantly influences overheads. EPC is a secure, SGX-exclusive, limited-size memory area that acts as a cache for encrypted enclave code and data. EPC enhances performance by minimizing costly paging to regular memory, which requires verification. Similarly, we observed higher performance without exposing the CPU core's second logical thread (hyperthread) to TDX. In its default configuration, PyTorch only executes on the first logical thread of a core, with hyperthreads introducing noise. We also identified more concerning limitations with non-uniform memory access (NUMA) and huge pages.

\subsubsection{Multiple sockets}




Figure~\ref{fig:two_NUMA} shows inference performance when deployed on two sockets. The performance overheads increase considerably, with TDX reporting an overhead of 12.11-23.81\%. There are two reasons for such performance. First, the socket interconnect has a dedicated cryptographic unit~\cite{johnson2024sgx}, and any data moving between sockets must be encrypted and integrity-protected, which incurs a performance penalty on the critical path. 

Second, TDX and SGX drivers lack working NUMA support. 
Figure~\ref{fig:70B} shows the performance of TDX when running on the 70B parameter model. This model is too large to fit into the memory of a single socket, and the 200ms service level is no longer upheld. We compare TDX performance to a VM with NUMA nodes bound in QEMU to the physical memory of two sockets (VM B) and non-bound (VM NB). While TDX is not as low-performing as VM NB, it has a considerable overhead compared to VM B, especially in terms of latency. We found that TDX's KVM driver does not adhere to the bindings that we provided.

We are not displaying the results of SGX as its overheads become prohibitively large, increasing up to 230\%. While encryption on the socket interconnect reduces performance, such performance in SGX is predominantly due to a lack of proper support for NUMA. The memory is presented to the application as a single unified NUMA node, 
potentially resulting in the allocation of all memory on a single socket. While efforts have been made to optimize allocations to align with the thread using the data~\cite{johnson2024sgx}, we have not found satisfactory performance of SGX in multiple sockets.

We also found that sub-NUMA clustering has a significant influence on both SGX and TDX. Sub-NUMA clustering (SNC)~\cite{mulnixIntelXeonProcessor} in Intel CPUs divides a single socket into multiple NUMA domains, aiming at improving performance for ML workloads. TEE drivers also do not support sub-NUMA domains, resulting in inefficient memory placement. 
In our test runs, using sub-NUMA domains increased overhead by more than eight times, from approximately 5\% to 42\%. As a result, we disabled sub-NUMA clustering.

\Insight{TDX and SGX do not properly support NUMA bindings, which leads to a considerably degraded performance, especially in the case of models that do not fit in the memory of a single socket.}

\subsubsection{Hugepages}

For TDX, we also identified that it does not use 1GB huge pages~\cite{panwarMakingHugePages2018}, which increases the number of necessary translation lookaside buffer (TLB) accesses, worsening memory access latency. Figure~\ref{fig:two_NUMA} also shows the performance of different VM hugepage allocation strategies. VM FH uses preallocated 1GB hugepages, and VM TH uses 2MB transparent hugepages. TDX overheads over VM TH remain the same order of magnitude as in the single socket case. We found that TDX in the background uses transparent huge pages even if 1GB pages are provided. A larger data movement in the case of two NUMA nodes implies greater TLB pressure, manifesting in larger overheads of VM TH and TDX compared to VM FH and bare metal, for which huge pages matter less. The overhead of VM TH over VM FH quantifies the performance cost due to the lack of 1 GB hugepage support in TDX at 3.19–5.20\%.

\Insight{TDX uses self-allocated transparent hugepages and ignores manually reserved hugepages, which costs up to 5\% of raw performance.}

\subsection{Per-block overheads}
\begin{figure}
  \centering
  \includegraphics[width=\linewidth]{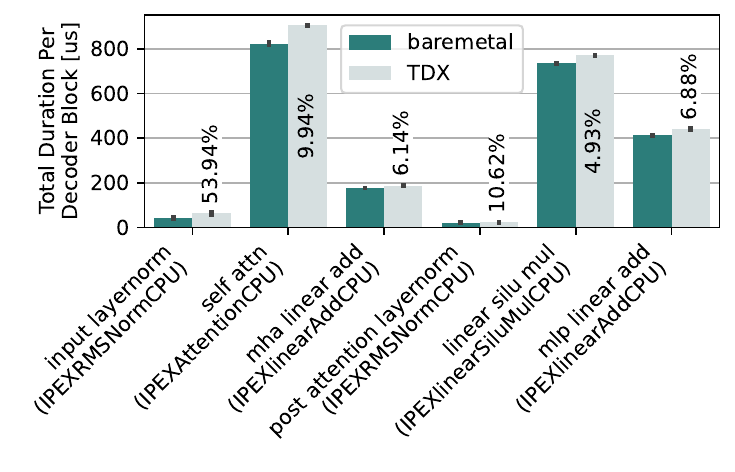}
  \vspace{-2em}
  \caption{The duration and TDX overhead of each decoder block layer for Llama7B on a single socket of EMR2.}
  \label{fig:layer_duration}
  \vspace{-1em}
\end{figure}

To better understand the sources of overhead, we traced the single-socket inference of 128 in/out tokens for a batch size of 4 for TDX.
We then parsed the traces to measure the time of each inference layer.
We observed that decoder blocks take 99.9\% of the time, with the remainder devoted to embedding and final normalization.
Figure~\ref{fig:layer_duration} shows the duration and overheads for each decoder block layer.
The most significant overheads are incurred in input and post-attention layer norms.
However, these have large relative noises and form only 3\% of the total block time.
The most considerable cost in raw performance is incurred in self-attention and linear SiLU multiplication.
Given that these have a considerable data movement~\cite{ivanovDataMovementAll2021}, it is clear that memory encryption is a major contributor to the overheads.
The time these take is impacted by the arithmetic intensity, influenced by solutions such as AMX and operational parameters such as batch and input sizes.
These two parameters also considerably impact the exact relative durations we have shown above.
As we increased the batch and input sizes, we observed that self-attention and linear SiLU remain the most significant contributors to overall block time, with self-attention dominating even more.
%

\begin{figure*}[t]
  \centering
  \includegraphics[width=\linewidth]{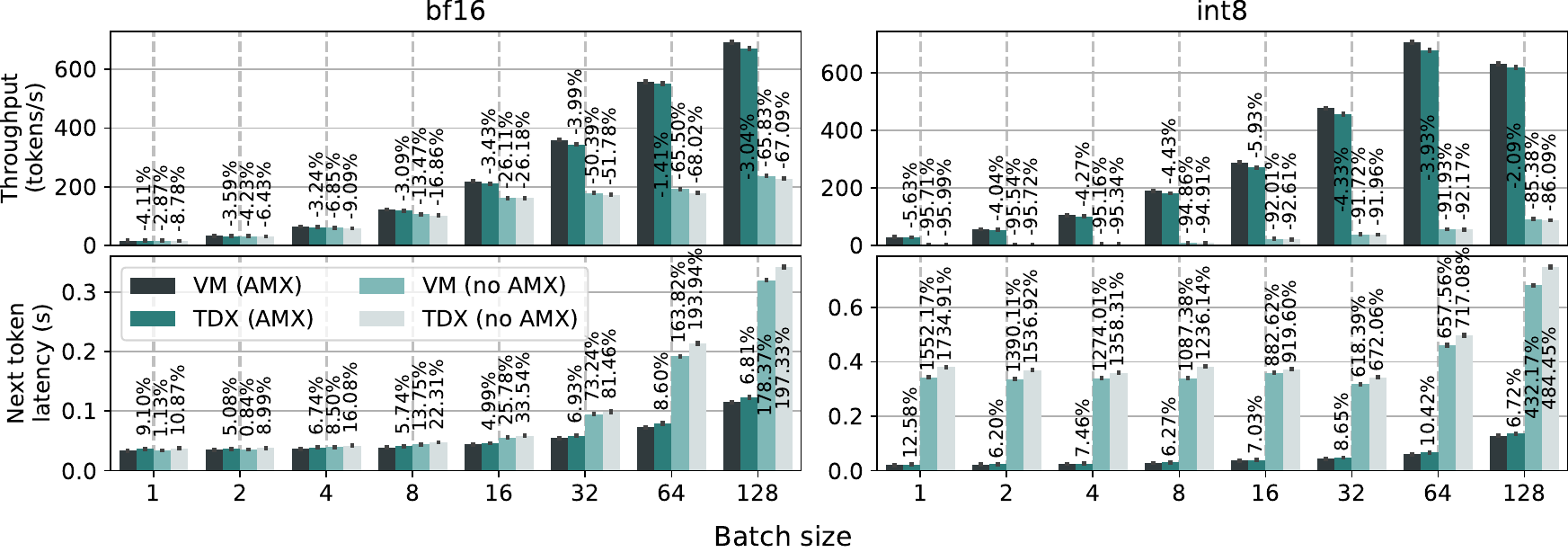}
  \caption{Comparison of performance between AMX and no-AMX systems as we scale the batch size for Llama2 7B, with 128 in and out tokens and beam size equal to one on EMR2. The overheads are relative to VM running AMX. We show the best performing setups: latency on two sockets, throughput on one.}
  \label{fig:AMX_scaling}
  \vspace{-1em}
\end{figure*}

\begin{figure*}[b]
  \centering
  \vspace{-1em}
  \includegraphics[width=\linewidth]{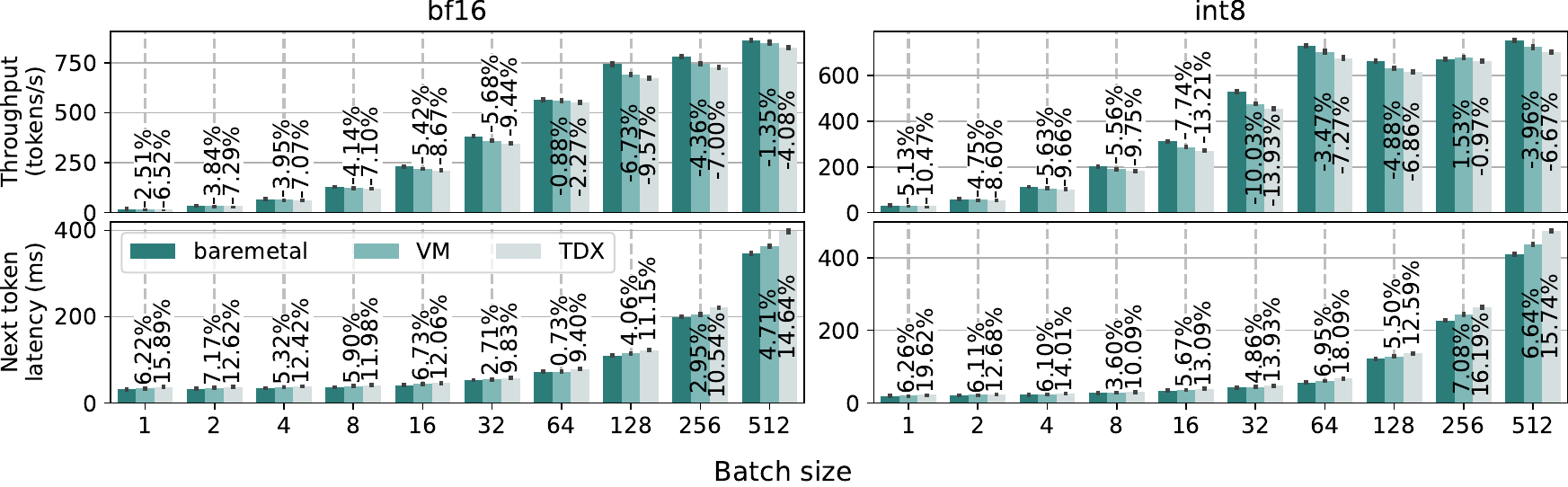}
  \caption{Comparison of next token latency and throughput as we scale the batch size with 128 in and out tokens and beam size of one on EMR2. Performance overheads are shown relative to bare metal. Latency is measured on two sockets, while throughput is measured on a single socket. Throughput for two sockets equals twice the shown values.}
  \label{fig:batch_scaling}
\end{figure*}

\subsection{Use of AMX}
\label{sec:AMX}
As shown in Section~\ref{sec:experimental_setup}, using IPEX, which leverages AMX, has a significant impact on inference performance. However, what we found is that AMX also minimizes TDX overheads. For further experiments, we focus solely on TDX, which performs worse than SGX, forming a lower bound on performance. However, it is easier to work with, especially for experiments that disable AMX, limit the number of cores, or run RAG pipelines. All VMs henceforth use 1GB hugepages.

Figure~\ref{fig:AMX_scaling} investigates the benefits of AMX across batch sizes, against a setup running IPEX without AMX. 
In the case of \verb|bfloat16|, AMX initially provides a slight advantage of 1-4\%, which increases to hundreds of percent with larger batch sizes (more compute). 
AMX not only significantly influences raw performance but also reduces the overheads of TDX, lowering them by up to 30\% for latency and up to 2\% for throughput. As latency results are measured on two sockets, lower NUMA traffic caused by AMX explains these benefits. Importantly, we also observed up to 96\% of overhead in throughput and 1700\% in latency for \verb|int8|. Such low performance occurs because the model quantization is fine-tuned for AMX, and there is a lack of AVX implementation for \verb|int8| in IPEX.

\Insight{AMX lowers TDX overheads, accelerates workloads up to 2.6x, and enables quantized inference.}

\subsection{Efficient use of memory}
\label{sec:optimizing_operational}

The final overheads we observed include memory protection costs, which are influenced by the amount of paging and the application's arithmetic intensity. We optimized the former by using TCMalloc~\cite{durnerImpactMemoryAllocation2019}, which reduces the memory pressure. For the latter, we used an OpenMP~\cite{dagumOpenMPIndustryStandard1998} version suitable for Intel processors. However, the choice of operational parameters, such as batch and input sizes, has a greater impact.

\begin{figure*}[b]
  \centering
  \vspace{-1em}
  \includegraphics[width=\linewidth]{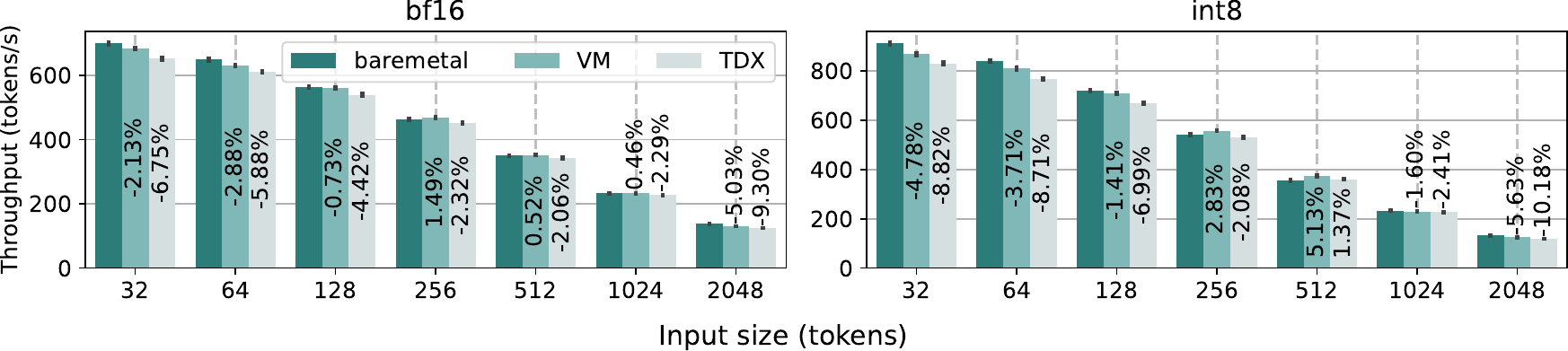}
  \caption{Comparison of generation throughput as we scale the input size for Llama2 7B on a single socket, with 128 out tokens, beam size 1, batch size 64, on EMR2. The overheads are relative to bare metal.}
  \label{fig:input_size_scaling}
\end{figure*}

\subsubsection{Batch size scaling}

Figure~\ref{fig:batch_scaling} shows the results of varying the batch size. As it is scaled, we expect more algorithmic intensity, which lowers TDX's overhead stemming from memory encryptions. This is precisely what we observe. For \verb|int8|, the workload saturates the throughput at batch size 64 when the overheads drop from 9-11\% to 6\% or less. \verb|bfloat16| also achieves throughput saturation, but around a batch size of 512. This is also when the overheads drop from 7-10\% to 4-7\%. From a latency perspective, we do not observe such a strong correlation, which is due to the overhead of socket interconnect data movement that also increases alongside algorithmic intensity. A batch size of 64 achieves the best performance for \verb|bfloat16| throughput, when the overheads drop to 2\%. As this marks the inflection point for bare metal performance, we evaluate it across different input sizes. 


\subsubsection{Input size scaling}
Figure~\ref{fig:input_size_scaling} shows the throughput performance against the input size. 
We observe that the overhead of TDX decreases as the input size increases, both for \verb|int8| and \verb|bfloat16|, until it reaches 2048 tokens. The overhead variability stems from the interplay of caches and AMX. As we initially increase the input size, we benefit from the workload saturating the AMX units and becoming more compute-bound, similarly to the batching case. However, as we increase the input size, the KV cache size per new token also grows. Eventually, it reaches the point where each token causes a considerable cache miss rate, making the workload memory-bound. We observe increased overheads for both TDX and VM, as this also leads to TLB misses. 
At this regime, we achieve overheads similar to smaller batch sizes.

\Insight{TDX has the lowest overhead when the workload is compute-bound.}

\section{GPU TEEs}
To put our CPU results in perspective, we also investigate cGPUs.
%
At the time of writing, the GPU-based TEE introduced by NVIDIA in the Hopper architecture is the only accelerated TEE solution entering the space on a large scale. 
H100s with CC enabled are available only in production mode at Azure~\cite{AnnouncingAzureConfidential} and GCP~\cite{PrivacypreservingConfidentialComputing}.
 Their successors, B100s, are currently not available in any CSP in the CC configuration.

\subsection{NVIDIA Confidential GPUs}
cGPUs require a host CPU TEE, enabling GPU attestation.
Users can run their kernels on cGPUs without any changes to existing CUDA applications.
All command buffers, kernels, and data transfers over PCIe between CPU and GPU are encrypted and authenticated via a bounce buffer.
This prevents hypervisor or physical attackers from accessing sensitive information.
These transfers, together with an additional kernel invocation latency, are the main costs of the current cGPUs.
To avoid the PCIe overhead, solutions such as PCIe IDE need to be used~\cite{IDETDISPOverview}.
While PCIe transfers are protected, the HBM memory of H100s is not.
Additionally, the NVLINK communication is unprotected when combining multiple H100s, requiring secure communication through the host.
%
The B100s resolve the main security issues of H100s and introduce HBM memory and NVLINK encryption. 
While B100s address these issues, their availability in CC configurations makes it challenging to evaluate the costs of these protections.

\subsection{Experimental setup}
We used an H100 NVL GPU with 94 GB of memory (\textasciitilde \$30,000~\cite{NVIDIAH100NVL}) rented from Azure (confidential \verb|NCCads_H100_v5| and non-confidential \verb|NCads_H100_v5|), with a 40-virtual CPU (vCPU) AMD EPYC 9V84 host and 320 GiB memory. We deployed Ubuntu 24.04 and leveraged vLLM~\cite{kwon2023efficient} version 0.8.5 as an optimized inference framework. As our machine is rented, we do not have access to bare metal and present the results for raw and Confidential GPUs (cGPU).

\begin{figure*}[t]
  \centering
  \includegraphics[width=\linewidth]{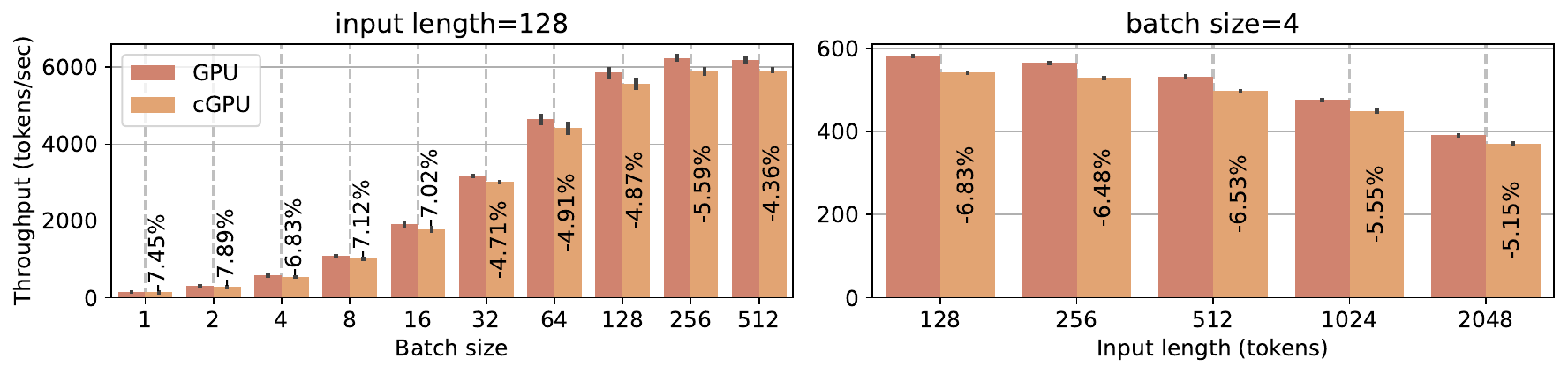}
  \caption{GPU throughput as a function of batch and input sizes. As both increase, the overheads are minimized, and oscillate between 7.5\% and 4.4\%}
  \label{fig:gpu_scaling}
  \vspace{-1em}
\end{figure*}

\subsection{Batch and input size scaling}
\label{sec:gpu_scaling}
Figure~\ref{fig:gpu_scaling} shows the performance of GPUs for Llama2 7B for different batch sizes and input lengths.
cGPUs exhibit similar performance to CPU TEEs, albeit with lower noise.
This is an expected behavior since GPUs do not have encrypted memory on the critical path. 
As both batch size and input length increase, the cGPU performance improves, primarily due to increased arithmetic intensity.
Since the share of time spent on setup remains roughly the same (including overhead-inducing kernel invocations and data transfers from the CPU), the overheads naturally decrease. 
While for inference, the data transfer is minimal, for workloads such as LLM training, it is large.
Solutions such as TDX Connect~\cite{chengIntelTDXDemystified2024} and SEV IO~\cite{SEVTIOFirmwareInterface2023} are in development to address these overheads.

\Insight{GPU TEEs achieve less than 10\% overheads, which decreases with larger batch and input sizes.}


\subsection{Comparing CPUs and GPUs}
\label{sec:gpu_perf}

\begin{figure*}[b]
  \centering
  \vspace{-1em}
  \includegraphics[width=\linewidth]{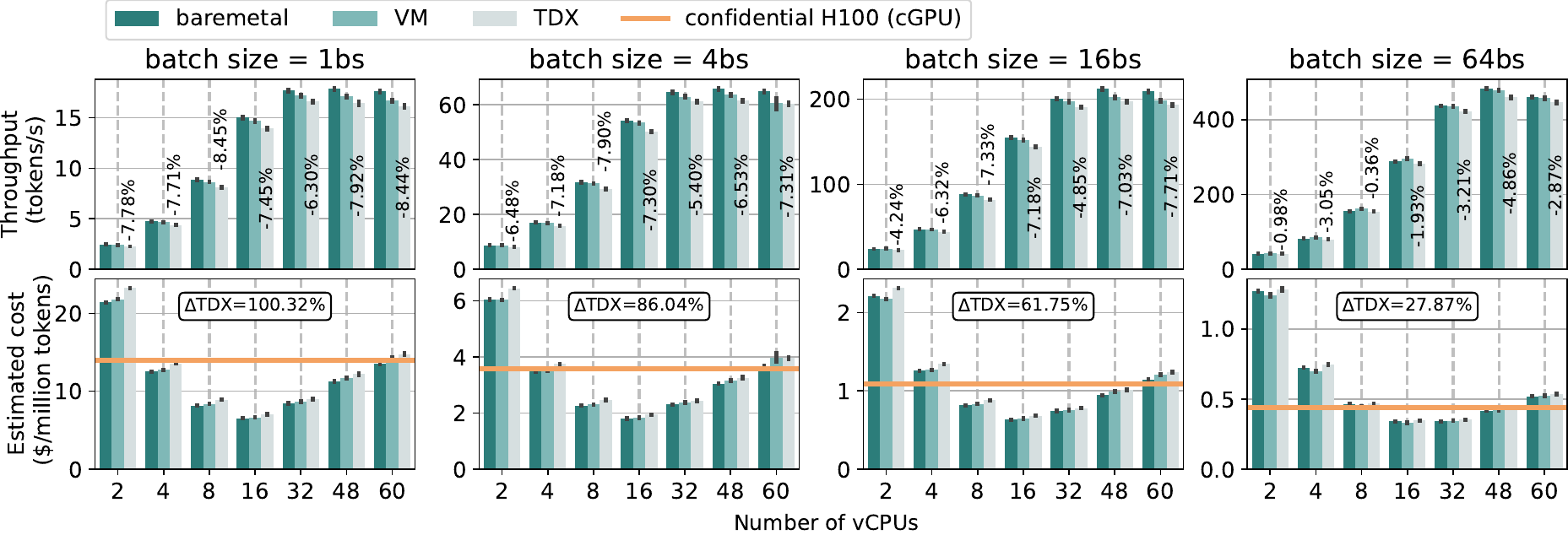}
  \caption{vCPU scaling and cost of generating on EMR2. Generation throughput includes the first token latency, measured over 128 in and out tokens on a single socket for bfloat16. The throughput overheads are with respect to bare metal, and the cost of overheads of TDX with respect to GPU.}
  \label{fig:vCPU_scaling_batches}
\end{figure*}

\begin{figure*}[t]
  \centering
  \includegraphics[width=\linewidth]{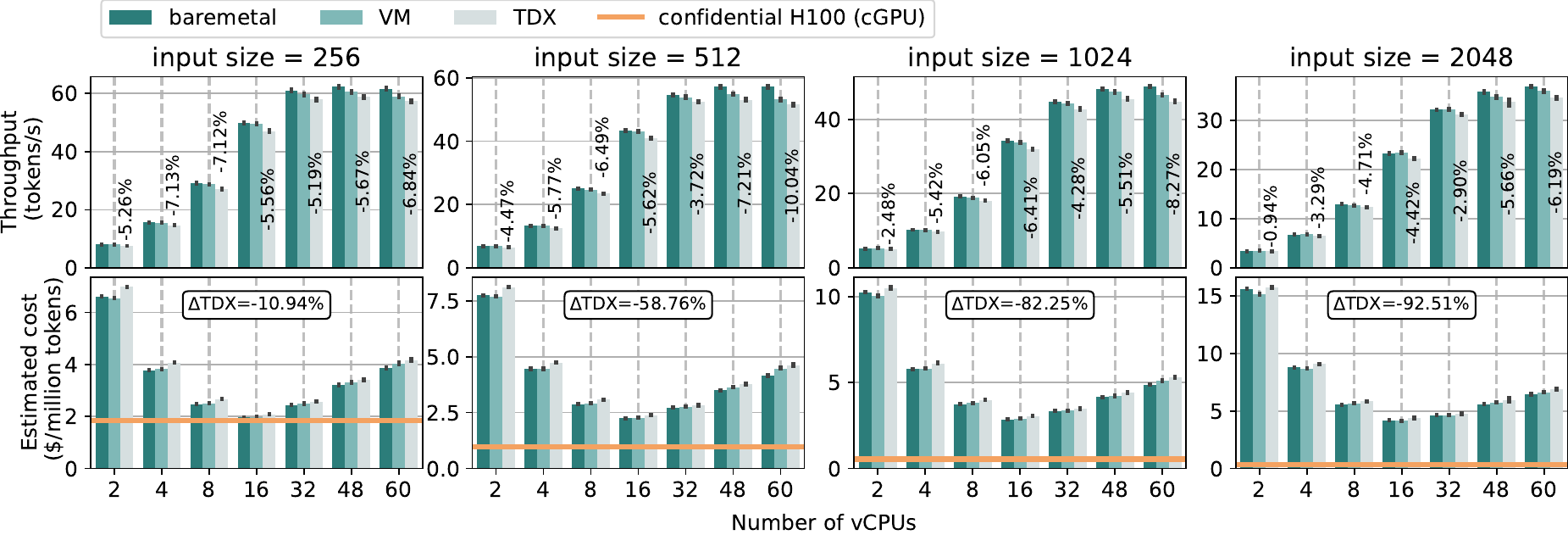}
  \caption{vCPU scaling and cost of generating tokens on EMR2. Throughput includes the first token latency, measured over 128 out tokens, batch size 4, on a single socket for bfloat16. The throughput overheads are with respect to bare metal, and the cost of cGPU overheads with respect to TDX.}
  \label{fig:vCPU_scaling_inputs}
  \vspace{-1em}
\end{figure*}

\subsubsection{Hybrid setups} Results in Section~\ref{sec:gpu_scaling} indicate that the GPU has a much better raw performance. This occurs as long as the model can be entirely fitted on the GPU. Prior research has shown~\cite{naUnderstandingPerformanceImplications2024} that if parts of the model need to be offloaded to the host memory, the AMX-accelerated CPUs outperform GPUs. This is even more so the case for confidential computing, as any data movement between CPU and GPU is more expensive, due to the cost of encrypting the PCIe bounce buffer. We demonstrate that in the case of confidential computing, two additional scenarios arise in which CPUs outperform current GPUs.
\subsubsection{Resource efficiency}
Figure~\ref{fig:vCPU_scaling_batches} shows the throughput across different batch sizes (columns) and numbers of CPU cores used during inference. The results indicate that the workload remains compute-bound until 32 cores, after which it becomes memory-bound, suggesting minimal performance gain above this number of cores. Similarly to prior plots, a batch size of 64 has the lowest TDX overheads. 

Additionally, Figure~\ref{fig:vCPU_scaling_batches} shows the cost of inference of 1 million tokens. To evaluate the cost of running different setups, we used spot prices offered by Google Cloud Platform (GCP) for the same machine type deployed in the US East 1 region. As GCP allows users to select the number of vCPU cores and the amount of memory separately, we assumed 128 GB of memory, which we found to be sufficient for deploying Llama2 7B in all the shown cases. We then scaled the number of vCPUs, keeping the memory size constant. Memory initially dominates the cost of renting, as it is fixed regardless of the number of CPU cores used. As we add cores, the performance increases, lowering the price per million tokens, which starts climbing back to 32 cores when the throughput plateau is reached. As we increase the batch size, the computational needs increase, making the larger machines more economical. For example, at a batch size of 128, 32 cores become optimal. As this workload becomes memory-bound easily, renting an almost 2x cheaper Sapphire Rapid performing up to 40\% worse~\cite{eadlineIntelWontHave2023}, provides an even more affordable alternative. 


We marked the cGPU cost-effectiveness with an orange line in Figure~\ref{fig:vCPU_scaling_batches}. While more performant, GPUs also have a significantly higher price per hour, resulting in cGPUs being up to 100\% more expensive. We observe that as the batch size increases, the advantage of CPU TEEs slowly fades, until it reaches a batch size of 128, at which point they equalize. Such behavior is expected as GPUs become more efficient with larger batch sizes~\cite{popeEfficientlyScalingTransformer2023}. LLM queries with low computational intensity are especially more cost-efficient when using TEEs. Currently, NVIDIA supports CC only on H100 and B100 systems, excluding MIG~\cite{liCharacterizingMultiInstanceGPU2022} and older or less powerful GPUs, such as the A100, which are used to optimize cost efficiency.


To verify that CPU TEEs eventually lose their advantage when compute requirements are sufficient, we also evaluate performance with varying input sizes for a batch size of 4. As results in Figure~\ref{fig:vCPU_scaling_inputs} reveal, from the cost perspective, CPU TEEs are considerably more sensitive to input size than cGPUs. 
The first batch size for which CPU TEEs are uncompetitive is 128, losing the 100\% resource advantage of batch size 1. However, we only needed to double the input size to achieve a similar reduction in gains, from 86\% to -10\%. As the attention part of the model grows quadratically with the input size, it implies a greater impact on compute requirements as compared to only linear increases for batch size.

\subsubsection{Security} While CPU TEEs perform worse than cGPUs with larger input sizes, they have one more advantage: security. CPU TEEs are more mature, and their security model is stricter than cGPUs. H100s do not encrypt their HBM memory~\cite{dhanuskodiCreatingFirstConfidential2023}, compared to CPUs that do. While in CPU-based systems, communication between different sockets is transparently encrypted, interconnects such as PCIe and NVLINK do not yet have this feature~\cite{dhanuskodiCreatingFirstConfidential2023}, which limits inter-accelerator communication to go through the host. 
This is crucial for larger models that do not fit on a single GPU.
While B100s address these issues, we expect that they will add a non-negligible overhead to H100s' results, since we identified memory encryption as a significant cost in CPUs. 


\subsubsection{Scaling models} We compared the CPUs (fitting ~70B parameters) to a single GPU (fitting ~30B parameters). Scale-up of confidential H100s is costly due to the aforementioned security concerns. Similarly, scaling out through combining single-GPU VMs is currently inefficient. As the cGPU instances do not support RDMA and GPUdirect, all data is transferred through the CPU, capping throughput at 3GB/s (considerably lower than the non-confidential 40GB/s)~\cite{yangDissectingPerformanceOverheads2025}. This is costly for throughput-hungry patterns such as pipeline parallelism and tensor parallelism. We expect this to lower the advantage of GPUs over CPUs. A network protection scheme, such as IPsec, is required on top of both CPUs and GPUs, which also introduces an overhead of up to 90\%~\cite{Chrapek2025SecPerf}.

\Insight{For strictest security workloads, and relatively small LLMs such as Llama2 7B, where H100 GPUs would be unsaturated (e.g., small batch or input sizes), CPU TEEs offer a pragmatic way to secure inference.}

\section{Moving to RAG}
\label{sec:RAG}

\begin{figure}[b]
  \vspace{-1em}
  \centering
  \includegraphics[width=\linewidth]{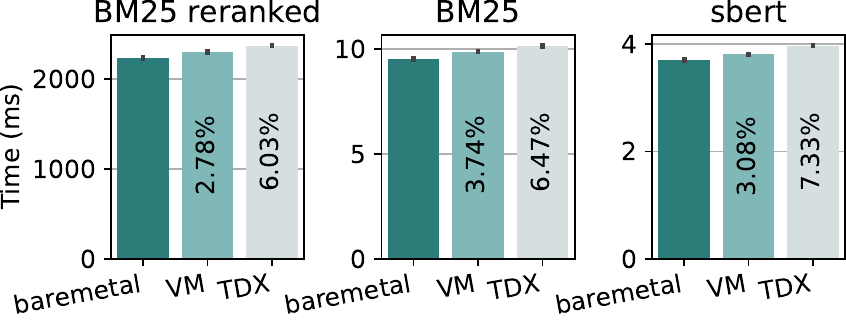}
  \caption{Comparison of mean evaluation time for RAG systems on EMR2.}
  \label{fig:rag}
\end{figure}

RAG is a practical showcase of our insights. RAG is an extension of LLMs, enabling them to retrieve documents that match queries. RAG embeds documents in an index, which is then searched during inference for closest matches in a process called retrieval. For example, the Best Matching 25 (BM25) is a classic retrieval model that ranks documents by keywords. Reranked BM25 first retrieves BM25 and then reranks it using a cross-encoder. For both, an Elasticsearch database~\cite {ElasticElasticsearch2025} is typically used to store the documents. RAG can also involve LLMs such as SBERT, which encodes queries and documents into dense vectors using a pre-trained Sentence-BERT encoder and ranks matches based on cosine similarity. We evaluate the performance of RAG using these three methods in BEIR~\cite{thakurBEIRHeterogeneousBenchmark2021}, running them and an Elasticsearch database entirely within TDX. Figure~\ref{fig:rag} shows that even though the RAG workload, such as BM25 ranking, differs from a normal LLM inference, our results display a similar level of overhead. We observe 6-7\% degradation for TDX, suggesting CPU TEEs might also be used for these purposes without significant performance impacts. Additionally, knowing that LLM RAG is conducted frequently with a batch size of one and for small models such as SBERT, we can leverage Insight 11 to deduce that CPU TEEs might be more cost-efficient than cGPUs.

\begin{table*}[]
\centering
\resizebox{\textwidth}{!}{%
\begin{tabular}{@{}rrrccc@{}}
\toprule
\multicolumn{3}{r}{System}                                                                                                                                       & Intel SGX (process TEE)                                                               & Intel TDX (VM TEE)                                                                        & H100 cGPU (GPU TEE)                                                      \\ \midrule
\multirow{5}{*}{\vspace{-1em}Security}                 & \multirow{2}{*}{Hardware}                                                                     & Memory               & \faBatteryFull                                                                        & \faBatteryFull                                                                            & \faBatteryEmpty~(HBM unencrypted)                                        \\
                                          &                                                                                               & Scale-up             & \faBatteryFull                                                                        & \faBatteryFull                                                                            & \faBatteryHalf~(NVLINK unprotected)                                      \\ \cmidrule(l){2-6} 
                                          & \multirow{3}{*}{Software}                                                                     & App                  & \faBatteryFull                                                                        & \faBatteryFull                                                                            & \faBatteryFull                                                           \\
                                          &                                                                                               & OS                   & \faBatteryHalf~(libOS)                                                                & \faBatteryFull                                                                            & \faBatteryFull                                                           \\
                                          &                                                                                               & VM                   & \faBatteryEmpty                                                                       & \faBatteryFull                                                                            & \faBatteryFull                                                           \\ \midrule
\multirow{6}{*}{\vspace{-3em}Performance} & Overhead                                                                                      & Single resource      & \textasciitilde4-5\%                                                                  & \textasciitilde5–10\%                                                                     & \textasciitilde4–8\%                                                     \\ \cmidrule(l){2-6} 
                                          & \multirow{4}{*}{\begin{tabular}[c]{@{}r@{}}Parameters\\ influencing\\ overheads\end{tabular}} & Batch size↑          & ↓                                                                                     & ↓                                                                                         & ↓                                                                        \\
                                          &                                                                                               & Input size↑          & ↓↑                                                                                    & ↓↑                                                                                        & ↓                                                                        \\
                                          &                                                                                               & AMX                  & ↓                                                                                     & ↓                                                                                         & -                                                                        \\
                                          &                                                                                               & Scale-up             & ↑↑                                                                                    & ↑                                                                                         & ↑↑                                                                       \\ \cmidrule(l){2-6} 
                                          & \begin{tabular}[c]{@{}r@{}}Sources\\ of overheads\end{tabular}                                &                      & \begin{tabular}[c]{@{}c@{}}EPC paging, \\ enclave exits, \\ memory, NUMA\end{tabular} & \begin{tabular}[c]{@{}c@{}}Virtualization tax, \\ hugepages, \\ memory, NUMA\end{tabular} & \begin{tabular}[c]{@{}c@{}}PCIe transfers, \\ kernel launch\end{tabular} \\ \midrule
\multirow{3}{*}{\vspace{-0.5em}Cost}        & Development                                                                                   &                      & \faBatteryHalf                                                                        & \faBatteryFull                                                                            & \faBatteryFull                                                           \\ \cmidrule(l){2-6} 
                                          & \multirow{2}{*}{\begin{tabular}[c]{@{}r@{}}Resource\\ efficiency\end{tabular}}                & Small inputs/batches & \faBatteryFull                                                                        & \faBatteryFull                                                                            & \faBatteryHalf                                                           \\
                                          &                                                                                               & Large inputs/batches & \faBatteryHalf                                                                        & \faBatteryHalf                                                                            & \faBatteryFull                                                           \\ \bottomrule
\end{tabular}%
}
\vspace{0.5em}
\caption{The summary of evaluated systems and the insights. \faBatteryFull\ indicates full/good, \faBatteryHalf\ partial, and \faBatteryEmpty\ no support. ↓ indicates decreasing, ↑ increasing, ↑↑ increasing considerably more than ↑, and ↓↑ first decreasing, then increasing overheads.}
\label{tab:summary}
\vspace{-2em}
\end{table*}


\Insight{Performance of entire RAG pipeline in TDX achieves similar overheads to the LLM inference.}




\section{Related work}
TEEs have been investigated in the past for protecting ML models~\cite{moMachineLearningConfidential2023}. Yet, most of these approaches offload only parts of the models to TEEs, providing weaker notions of security and citing low TEE performance as the reason. For example, Slalom~\cite{tramerSlalomFastVerifiable2019} offloads linear layers to the GPU with a probabilistic algorithm guaranteeing some security. Furthermore, none of these works explored LLMs, focusing instead on simpler models due to the extensive model changes. 
In contrast, we run an entire LLM inference pipeline in TEEs, demonstrating their practicality for protecting LLMs. 

Some performance studies have been conducted on SGX~\cite{miwaAnalyzingPerformanceImpact2023a, zhaoPerformanceIntelSGX2016, dinhngocEverythingYouShould2019a, gjerdrumPerformanceTrustedComputing2017, akramPerformanceAnalysisScientific2021} and TDX~\cite{misonoConfidentialVMsExplained2024, ExperimentalEvaluationTEE2025}. These focus on quantifying the overheads of the underlying primitive operations, such as memory overheads, and the performance of certain applications. However, none address workloads as compute-intensive as LLMs. Some works that demonstrate secure LLM inference~\cite{chrapekFortifyYourFoundations2024} focus more on security, missing the depth and key deployment insights, such as AMX performance improvements, scalability, and cost considerations. Similarly, GPUs have been studied for their sources of overhead~\cite{mohanSecuringAIInference2024, yangDissectingPerformanceOverheads2025}. However, these outline overheads considerably larger than ours, or do not show raw LLM performance. 
Additionally, none compares GPU TEEs to CPU TEEs, thereby failing to display the full spectrum of practical deployments.

\section{Conclusions}
We investigated several methods for protecting LLM deployments and discussed how TEEs yield a practical balance between security, performance, and programmability. We demonstrated the viability of securing LLMs with TEEs by running an inference pipeline on top of Intel's TDX and SGX, as well as NVIDIA's H100s. We conducted a thorough study of the performance of TEEs in these workloads, identifying the best frameworks, sources of overheads, and optimal operating points. We shared 12 key insights, showing, among others, that CPU TEEs have NUMA and hugepages issues, and how AMX helps improve their performance. We have also compared CPU and GPU TEEs in terms of performance, cost-efficiency, and security. Finally, we applied our lessons to a RAG pipeline within a TEE, demonstrating its performance. Table~\ref{tab:summary} shows the summary of our investigation. Our results show that TEEs impose a manageable performance overhead on LLM pipelines, demonstrating that TEEs represent a viable solution for protecting LLM inference, positioning them as a cornerstone for future confidential AI deployments.

\section*{Acknowledgment}

This research was conducted as part of the “UrbanTwin: An
urban digital twin for climate action: Assessing policies
and solutions for energy, water and infrastructure” project,
funded by ETH-Domain Joint Initiative program in the
Strategic Area Energy, Climate and Sustainable Environment, with additional support from Intel Corporation. We
thank Intel for providing hardware resources, Cory Cornelius, Anjo Vahldiek-Oberwagner, Marcin Spoczynski, Scott Constable and Mona Vij for their valuable feedback, and Madlen Koblinger for assisting with the design of figures.

\bibliographystyle{IEEEtranS}
\bibliography{library}

\end{document}